# MAP/REDUCE DESIGN AND IMPLEMENTATION OF APRIORIALGORITHM FOR HANDLING VOLUMINOUS DATA-SETS

Anjan K Koundinya[1],Srinath N K[1],K A K Sharma[1], Kiran Kumar[1], Madhu M N[1] and Kiran U Shanbag[1]

[1]Department of Computer Science and Engineering,
R V College of Engineering, Bangalore, India
annjank2@gmail.com

## ABSTRACT

*Apriori is one of the key algorithms to generate frequent itemsets. Analysing frequent itemset is a crucial step in analysing structured data and in finding association relationship between items. This stands as an elementary foundation to supervised learning, which encompasses classifier and feature extraction methods. Applying this algorithm is crucial to understand the behaviour of structured data. Most of the structured data in scientific domain are voluminous. Processing such kind of data requires state of the art computing machines. Setting up such an infrastructure is expensive. Hence a distributed environment such as a clustered setup is employed for tackling such scenarios. Apache Hadoop distribution is one of the cluster frameworks in distributed environment that helps by distributing voluminous data across a number of nodes in the framework. This paper focuses on map/reduce design and implementation of Apriori algorithm for structured data analysis.*

## 1. INTRODUCTION

In many applications of the real world, generated data is of great concern to the stakeholder as it delivers meaningful information / knowledge that assists in making predictive analysis. This knowledge helps in modifying certain decision parameters of the application that changes the overall outcome of a business process. The volume of data, collectively called data-sets, generated by the application is very large. So, there is a need of processing large data-sets efficiently. The data-set collected may be from heterogeneous sources and may be structured or unstructured data. Processing such data generates useful patterns from which knowledge can be extracted. the simplest approach is to use this template and insert headings and text into it as appropriate.

Data mining is the process of finding correlations or patterns among fields in large data-sets and building up the knowledge-base, based on the given constraints. The overall goal of data mining is to extract knowledge from an existing data-set and transform it into a human-understandable structure for further use. This process is often referred to as Knowledge Discovery in data-sets (KDD). The process has revolutionized the approach of solving the complex real-world problems. KDD process consists of series of tasks like selection, pre-processing, transformation, data mining and interpretation as shown in Figure1.

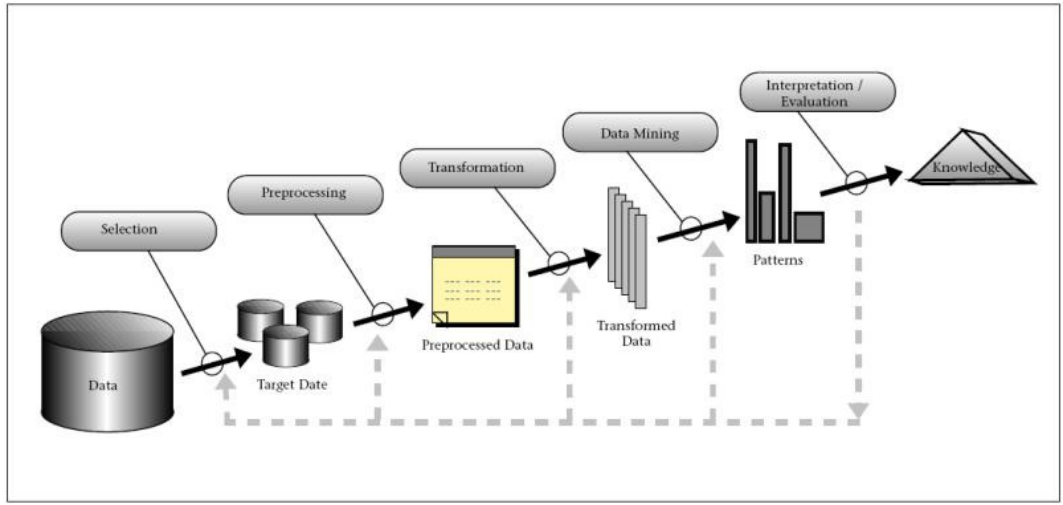

Figure 1: KDD Process

In a distributed computing environment is a bunch of loosely coupled processing nodes connected by network. Each nodes contributes to the execution or distribution / replication of data. It is referred to as a cluster of nodes. There are various methods of setting up a cluster, one of which is usually referred to as cluster framework. Such frameworks enforce the setting up processing and replication nodes for data. Examples are DryAdLinq and Apache Hadoop (also called Map / Reduce). The other methods involve setting up of cluster nodes on ad-hoc basis and not being bound by a rigid framework. Such methods just involve a set of API calls basically for remote method invocation (RMI) as a part of inter-process communication. Examples are Message Passing Interface (MPI) and a variant of MPI called MPJExpress.

The method of setting up a cluster depends upon the data densities and up on the scenarios listed below:

- The data is generated at various locations and needs to be accessed locally most of the time for processing.

- The data and processing is distributed to the machines in the cluster to reduce the impact of any particular machine being overloaded that damages its processing

This paper is organized as follows, the next section will discuss about complete survey of the related work carried out the domain of the distributed data mining, specially focused on finding frequent item sets. The section 3 of this paper discusses about the design and implementation of the Apriori algorithm tuned to the distributed environment, keeping a key focus on the experimental test-bed requirement. The section 4, discusses about the results of the test setup based on Map/Reduce – Hadoop. Finally conclude our work with the section 5.

## 2. RELATED WORK

Distributed Data Mining in Peer-to-Peer Networks (P2P) [1] offers an overview of distributed data-mining applications and algorithms for peer-to-peer environments. It describes both exact and approximate distributed data-mining algorithms that work in a decentralized manner. It illustrates these approaches for the problem of computing and monitoring clusters in the data residing at the different nodes of a peer-to-peer network. This paper focuses on an emerging branch of distributed data mining called peer-to-peer data mining. It also offers a sample of exact and approximate P2P algorithms for clustering in such distributed environments.

Web Service-based approach for data mining in distributed environments [2] presents an approach to develop a data mining system in distributed environments. This paper presents a web service-based approach to solve these problems. The system is built using this approach offers a uniform presentation and storage mechanism, platform independent interface, and a dynamically extensible architecture. The proposed approach in this paper allows users to classify new incoming data by selecting one of the previously learnt models.

Architecture for data mining in distributed environments [3] describes system architecture for scalable and portable distributed data mining applications. This approach presents a document metaphor called \emph{Living Documents} for accessing and searching for digital documents in modern distributed information systems. The paper describes a corpus linguistic analysis of large text corpora based on colocations with the aim of extracting semantic relations from unstructured text.

Distributed Data Mining of Large Classifier Ensembles [4] presents a new classifier combination strategy that scales up efficiently and achieves both high predictive accuracy and tractability of problems with high complexity. It induces a global model by learning from the averages of the local classifiers output. The effective combination of large number of classifiers is achieved this way.

Multi Agent-Based Distributed Data Mining [5] is the integration of multi-agent system and distributed data mining (MADM), also known as multi-agent based distributed data mining. The perspective here is in terms of significance, system overview, existing systems, and research trends. This paper presents an overview of MADM systems that are prominently in use. It also defines the common components between systems and gives a description of their strategies and architecture.

Preserving Privacy and Sharing the Data in Distributed Environment using Cryptographic Technique on Perturbed data [6] proposes a framework that allows systematic transformation of original data using randomized data perturbation technique. The modified data is then submitted to the system through cryptographic approach. This technique is applicable in distributed environments where each data owner has his own data and wants to share this with the other data owners. At the same time, this data owner wants to preserve the privacy of sensitive data in the records.

Distributed anonymous data perturbation method for privacy-preserving data mining [7] discusses a light-weight anonymous data perturbation method for efficient privacy preserving in distributed data mining. Two protocols are proposed to address these constraints and to protect data statistics and the randomization process against collusion attacks.An Algorithm for Frequent Pattern Mining Based on Apriori[8] proposes three different frequent pattern mining approaches (Record filter, Intersection and the Proposed Algorithm) based on classical Apriori algorithm. This paper performs a comparative study of all three approaches on a data-set of 2000 transactions. This paper surveys the list of existing association rule mining techniques and compares the algorithms with their modified approach.

Using Apriori-like algorithms for Spatio-Temporal Pattern Queries [9] presents a way to construct Apriori-like algorithms for mining spatio-temporal patterns. This paper addresses problems of the different types of comparing functions that can be used to mine frequent patterns.Map-Reduce for Machine Learning on Multi core [10] discusses ways to develop a broadly applicable parallel programming paradigm that is applicable to different learning algorithms. By taking advantage of the summation form in a map-reduce framework, this paper tries to parallelize a wide range of machine learning algorithms and achieve a significant speed-up on a dual processor cores.

Using Spot Instances for MapReduce Work flows [11] describes new techniques to improve the runtime of MapReducejobs. This paper presents Spot Instances (SI) as a means of attaining performance gains at low monetary cost.

## 3. DESIGN AND IMPLEMENTATION

### 3.1 Experimental Setup

The experimental setup has three nodes connected to managed switch linked to private LAN. One of these nodes is configured as Hadoop Master or as the namenode which controls the data distribution over the Hadoop cluster. All the nodes are identical in terms of the system configuration i.e., all the nodes have identical processor - Intel Core2 Duo and assembled by standard manufacturer. As investigative effort,configuration made to understand Hadoop will have three nodes in fully distributed mode. The intention is to scale the number of nodes by using standard cluster management software that can easily add new nodes to Hadoop rather than installing Hadoop in every node. The visualization of this setup is shown in the figure 2.

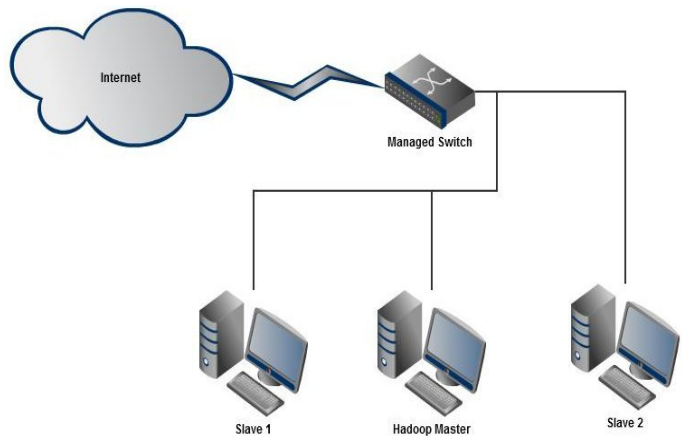

Figure 2: Experimental Setup for Hadoop Multi-node

### 3.1.1 Special Configuration Setup for Hadoop

Focus here is on Hadoop setup with multi-node configuration. The prerequisite for Hadoop installation is mentioned below -

- Set-up requires Java installed on the system, before installing Hadoop make sure that the JDK is up and running.
- The operating system can either be Ubuntu or Suse or Fedora for the setup.
- Create a new user group and user on Linux system. This has to be replicated for all the nodes planned for cluster setup. This user and group will exclusively usedfor Hadoop operations.

### 3.1.1.1 Single Node Setup

Hadoop must first be configured to the single node setup also called as Pseudo-distributed mode which runs on standalone machine to start with. Later this setup will be extended to fully –configured multi-node setup with al-least two nodes and with basic network interconnect as Cat 6 cable.  Below is the required configuration setup for single node setup-

- Download Hadoop version 0.20.x from Apache Hadoop site, which is downloaded as tar ball (tar.gz). Change the ownership of the file as 755 or 765 (octal representation for R W E).
- Unzip the File to /usr/local and create a folder Hadoop.
- Update the bashrc file located at \$HOME/.bashrc. Content to be appended to this file is illustrated at the *http://www.michael-noll.com/tutorials/running-hadoop-on-ubuntu-linux-single-node-cluster/#update-home-bashrc*.
- Modify the shell script - **hadoop-env.sh** and the contents to be appended is described at the *http://www.michael-noll.com/tutorials/running-hadoop-on-ubuntu-linux-single-node-cluster/#hadoop-env-sh*.
- Change the configuration file **conf/core-site.xml**, **conf/mapred-site.xml** and **conf/hdfs-site.xml** with the contents illustrated at the *http://www.michael-noll.com/tutorials/running-hadoop-on-ubuntu-linux-single-node-cluster/#conf-site-xml*.
- Format the HDFS file system via Namenode by using the command given below –

    ***$ hduser@ubuntu: $ /usr/local/hadoop/bin/hadoopnamenode–format***

- Start a single node cluster using the command -

    ***$ hduser@ubuntu: $ /usr/local/hadoop/bin/start-all.sh***

    This will start up a Namenode, Datanode, Jobtracker and a Tasktracker.

- Confgure the SSH access for the user created for Hadoop on the Linux machine.

### 3.1.1.2 Multi-Node Configuration

- Assign names to nodes by editing the files "/etc/hosts" . For instance -

***$ sudo vi /etc/hosts***

**192.168.0.1 master**
**192.168.0.2 slave1**
**192.168.0.3 slave2**

- SSH access -The hduser on master node must be able to connect to its own user account and also hduser on slave nodes through password-less SSH login. To connect the master to itself -

**hduser@master: $ ssh master**

- Configuration for Master Node -

    o **conf/masters** -This file specifies on which machine Hadoop starts secondary NameNodes in multi-node cluster. bin/start-dfs.sh and bin/start-mapred.sh shell scripts will run the primary NameNode and JobTracker on which node runs these commands respectively. Now update the file conf/masters like this

    o **conf/slaves** - This File actually consists of the host machines where Hadoop slave daemonsare run. Update the file by adding these lines -

**master**
**slave**

- Modify the following configuration files previously addressed under Single node setup-

    o In the file**conf/core-site.xml**

            **<property >**
            **<name>fs.default.name </name>**
            **<value>hdfs://master:54310 </value>**
            **<description>The name of the**
            **default file system.A URI whose schemeand authority determine**
            **the FileSystemimplementation.Theuri's scheme**
            **determines the con_g property(fs.SCHEME.impl)**
            **naming theFileSystem implementation class.Theuri's authority is usedto determine the host,**
            **port, etc. for a filesystem.**
            **</description >**
            **</property >**

    o In the file **conf/mapred-site.xml**
            **<property >**
            **<name>mapred.job.tracker</name>**
            **<value>master:54311 </value>**
            **<description>The host and port that the MapReduce job tracker runs at. If "local", then jobs are run in-process as a single map and reduce task.**

        **</description >**
        **</property >**

- o   In the file **conf/hdfs-site.xml**

        <property >
        <name>dfs.replication</name>
        <value>3 </value>
        <description>Default block replication. The actual number of replications
        can be specified when the file is created.The default is used if replication
        is not specified in create time.
        </description >
        </property >

- Format the HDFS via Namenode

        **/bin/start-dfs.sh**

This command will brings up the HDFS with NameNode on and the DataNodes on machines listed in /conf/slaves-file.

### 3.1.1.3 Running the Map/Reduce Daemon

To run mapred daemons, we need to run the following command on ?master?

hduser@master:/usr/local/hadoop\$ bin/start-mapred.sh

See if all daemons have been started or not by running the command jps. i.e. on master node

hduser@master:/usr/local/hadoop\$ jps

To stop all the daemons, we have to run this command/bin/stop-all.sh.

### 3.1.1.4 Running the Map-Reduce job

- To run code on Hadoop cluster, put the code into a jar file and name it as apriori.jar. Put input files in the directory /usr/hduser/input and output files in /usr/hduser/output/.

- Now in order to run the jar file and run the following command at HADOOP\_HOME prompt \\hduser@master:/usr/local/hadoop\$ bin/hadoop jar apriori.jar apriori /user/hduser/input /user/hduser/output\\

### 3.2 System Deployment

The overall deployment of the required system is visualized using the system organization as described in the figure 3. Series of Map calls is made to send the data to cluster node and the format is of the form <Key, Value> and then a Reduce calls is applied to summarize the resultant from different nodes. A simple GUI is sufficient to display these results to user operating the Hadoop Master.

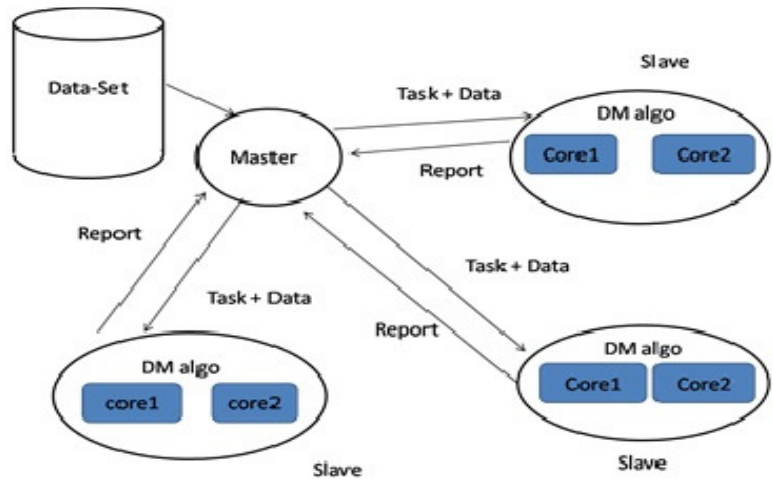

Figure 3: System Organization

### 3.3 Algorithm Design

The algorithm discussed produces all the subsets that would be generated from the given Item set. Further these subsets are searched against the data-sets and the frequency is noted.There are lots of data items and their subsets, hence they need to be searched them simultaneously so that search time reduces. Hence, the Map-Reduce concept of the Hadoop architecture comes into picture.Map function is forked for every subset of the items. These maps can run on any node in the distributed environment configured under Hadoop configuration. The job distribution is taken care by the Hadoop system and the files, data-sets required are put into HDFS. In each Map function, the value is the item set. The whole of the data-set is scanned to find the entry of the value item set and the frequency is noted. This is given as an output to the Reduce function in the Reduce class defined in the Hadoop core package.

In the reducer function, each output of the each map is collected and it is put into required file with its frequency. Algorithm is mentioned below in natural language:

- Read the subsets file.

- For each subsets in the file

    o Initialise count to zero and invoke map function.

    o Read the database.

    o For each entry in the database

        ▪ Compare with the subset.

- ▪ If equals the increment the count by one.
- ▪ Write the output counts into the intermediate file.
- • Call reduce function to count the subsets and report the output to the file.

## 4. RESULTS

The experimental setup described before has been rigorous tested against a Pseudo-distributed configuration of Hadoop and with standalone PC for varying intensity of data and transaction. The fully configured multi-node Hadoop with differential system configuration (FHDSC) would take comparatively long time to process data as against the fully configured similar multi-nodes (FHSSC)). Similarity is in terms of the system configuration starting from computer architecture to operating system running in it. This is clearly depicted in the figure 4.

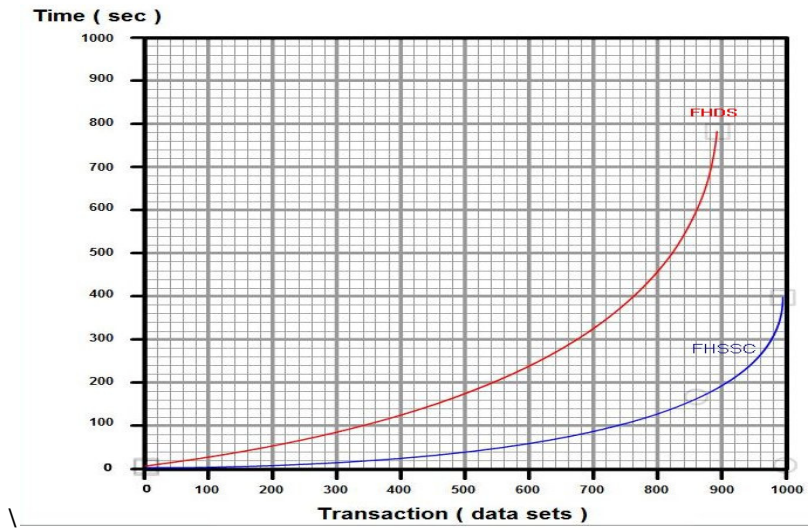

Figure 4:FHDSC Vs. FHSSC

The results for taken from the 3-node Fully-distributed and Pseudo distributed modes of Hadoop for large transaction are fairly good till it reaches the maximum threshold capacity of nodes. The result is depicted in the figure 5.

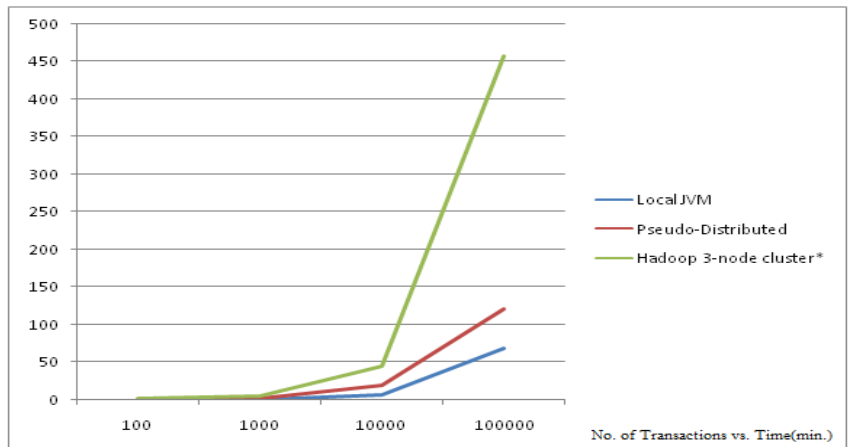

Figure 5:Transactions VsHadoop Configuration

Looking the graph, there is large variance in time seen at threshold of 12,000 transactions. Beyond which the time is in exponential. This is because of the computer architecture and limited storage capacity of 80GB per Node. Hence the superset transaction generation will take longer time to compute and the miner for frequent item-set.

The performance is expressed in the lateral comparison between FHDSC and FHSSC as given below-

$$\eta = \frac{FHDSC}{FHSSC}$$

$$\textbf{FHDSC = FHSSC} = \log_e N$$

Where N is the number of nodes installed in the cluster.

## 5. CONCLUSIONS AND FUTURE ENHANCEMENTS

The paper presents a novel approach of design algorithms for clustered environment. This is applicable to scenarios when there data-intensive computation is required. Such setup give a broad avenue for investigation and research in Data mining. Looking the demand for such algorithm there is urgent need to focus and explore more about clustered environment specially for this domain.

The future works concentrates about building a framework that deals with unstructured data analysis and complex Data mining algorithms. Also to look at integrating one of the following parallel computing infrastructure with Map/Reduce to utilize computing power for complex calculation on data especially in case of the classifiers under supervised learning-

- MPJ express is Java based parallel computing toolkit that can be integrated with distributed environment.
- AtejiPX  parallel programming toolkit with Hadoop.

## ACKNOWLEDGEMENTS

Prof. Anjan K would like to thanks Late Dr. V.K Ananthashayana, Erstwhile Head, Department of Computer Science and Engineering, M.S.Ramaiah Institute of Technology, Bangalore, for igniting the passion for research.Authors would also like to thank Late. Prof. Harish G for his consistent effort and encouragement throughout this research project.

## REFERENCES

[1]SouptikDatta, KanishkaBhaduri, Chris Giannella, Ran Wolff, and HillolKargupta, Distributed Data Mining in Peer-to-Peer Networks, Universityof Maryland, Baltimore County, Baltimore, MD, USA, Journal IEEEInternet Computing archive Volume 10 Issue 4, Pages 18 - 26, July 2006.

[2]Ning Chen, Nuno C. Marques, and NarasimhaBolloju, A Web Servicebasedapproach for data mining in distributed environments, Departmentof Information Systems, City University of Hong Kong, 2005.

[3] MafruzZamanAshrafi, David Taniar, and Kate A. Smith, A Data MiningArchitecture for Distributed Environments, pages 27-34, Springer-VerlagLondon, UK, 2007.

[4] GrigoriosTsoumakas and IoannisVlahavas, Distributed Data Mining ofLarge Classifier Ensembles, SETN-2008, Thessaloniki, Greece, Proceedings,Companion Volume, pp. 249-256, 11-12 April 2008.

[5] VudaSreenivasaRao, Multi Agent-Based Distributed Data Mining: AnOver View, International Journal of Reviews in computing, pages 83-92,2009.

[6] P.Kamakshi, A.VinayaBabu, Preserving Privacy and Sharing the Datain Distributed Environment using Cryptographic Technique on Perturbeddata, Journal Of Computing, Volume 2, Issue 4, ISSN 21519617, April2010.

[7] Feng LI, Jin MA, Jian-hua LI, Distributed anonymous data perturbationmethod for privacy-preserving data mining, Journal of Zhejiang UniversitySCIENCE A ISSN 1862-1775, pages 952-963, 2008.
[8] Goswami D.N. et. al., An Algorithm for Frequent Pattern Mining BasedOn Apriori (IJCSE) International Journal on Computer Science andEngineering Vol. 02, No. 04, 942-947, 2010.

[9] MarcinGorawski and PawelJureczek, Using Apriori-like Algorithmsfor Spatio-Temporal Pattern Queries, Silesian University of Technology,Institute of Computer Science, Akademicka 16, Poland, 2010.

[10] Cheng-Tao Chu et. al., Map-Reduce for Machine Learning on Multicore,CS Department, Stanford University, Stanford, CA, 2006.

[11] NavrajChohanet. al., See Spot Run: Using Spot Instances for Map-Reduce Workflows, Computer Science Department, University of California,2005.

# AUTHORS PROFILE

Anjan K Koundinya has received his B.E degree from Visveswariah Technological University, Belgaum, India in 2007 And his master degree from Department of Computer Science and Engineering, M.S. Ramaiah Institute of Technology, Bangalore, India. He has been awarded Best Performer PG 2010 and rank holder for his academic excellence. His areas of research includes Network Security and Cryptology, Adhoc Networks, Mobile Computing, Agile Software Engineering and Advanced Computing Infrastructure. He is currently working as Assistant Professor in Dept. of Computer Science and Engineering, R V College of Engineering.

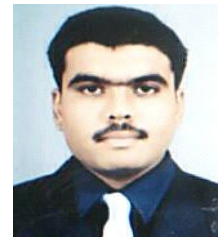

Srinath N K has his M.E degree in Systems Engineering and Operations Research from Roorkee University, in 1986 and PhD degree from AvinashLingum University, India in 2009. His areas of research interests include Operations Research, Parallel and Distributed Computing, DBMS, Microprocessor.His is working as Professor and Head, Dept of Computer Science and Engineering, R V College of Engineering.

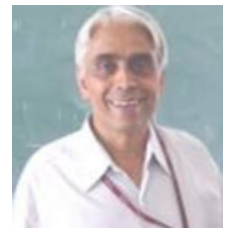